\documentclass[11pt]{article}
\usepackage[final]{acl}

\usepackage{times}
\usepackage{latexsym}
\usepackage{hyperref}
\hypersetup{
    colorlinks=true,
    linkcolor=blue,
    filecolor=magenta,      
    urlcolor=blue, 
}
\usepackage[T1]{fontenc}

\usepackage[utf8]{inputenc}

\usepackage{microtype}

\usepackage{inconsolata}

\usepackage{graphicx}
\graphicspath{{../figures/}{figures/}}
\usepackage{xcolor}

\title{Runtime Skill Audit: Targeted Runtime Probing for Agent Skill Security}

\author{Tu Lan\\
  Johns Hopkins University\\
  \And
  Chaowei Xiao \\
  Johns Hopkins University \\
  }

\begin{document}
\maketitle
\begin{abstract}
Agent skills let LLM agents reuse instructions, resources, tools, and workflows, but they also create a new place for malicious behavior to hide. A skill may look benign in its documentation or code while becoming harmful only when it is invoked with particular user requests, local assets, persistent state, or multi-step tool interactions. This makes purely static vetting brittle. We present Runtime Skill Audit (RSA), a dynamic analysis method that audits skills by asking what the skill-mediated agent actually does under targeted runtime conditions. Instead of testing every skill with the same generic tasks, RSA profiles risk-relevant interfaces, prepares the execution context needed to exercise them, and assigns security labels from the resulting trace evidence. We instantiate RSA on OpenClaw and evaluate it on 100 skills against representative static baselines. RSA achieves 90.0\% accuracy with an 88.0\% true positive rate and an 8.0\% false positive rate, improving accuracy by 13.0 percentage points over the best static baseline. Under self-evolving attacks, static detectors collapse after one or two rounds, while RSA continues to detect 19--20 out of 20 malicious skills across rounds.Our code is available at \href{https://github.com/tu-tuing/Runtime-Skill-Audit}{https://github.com/tu-tuing/Runtime-Skill-Audit}.
\end{abstract}
\begin{figure}[t]
    \centering
    \includegraphics[width=\columnwidth]{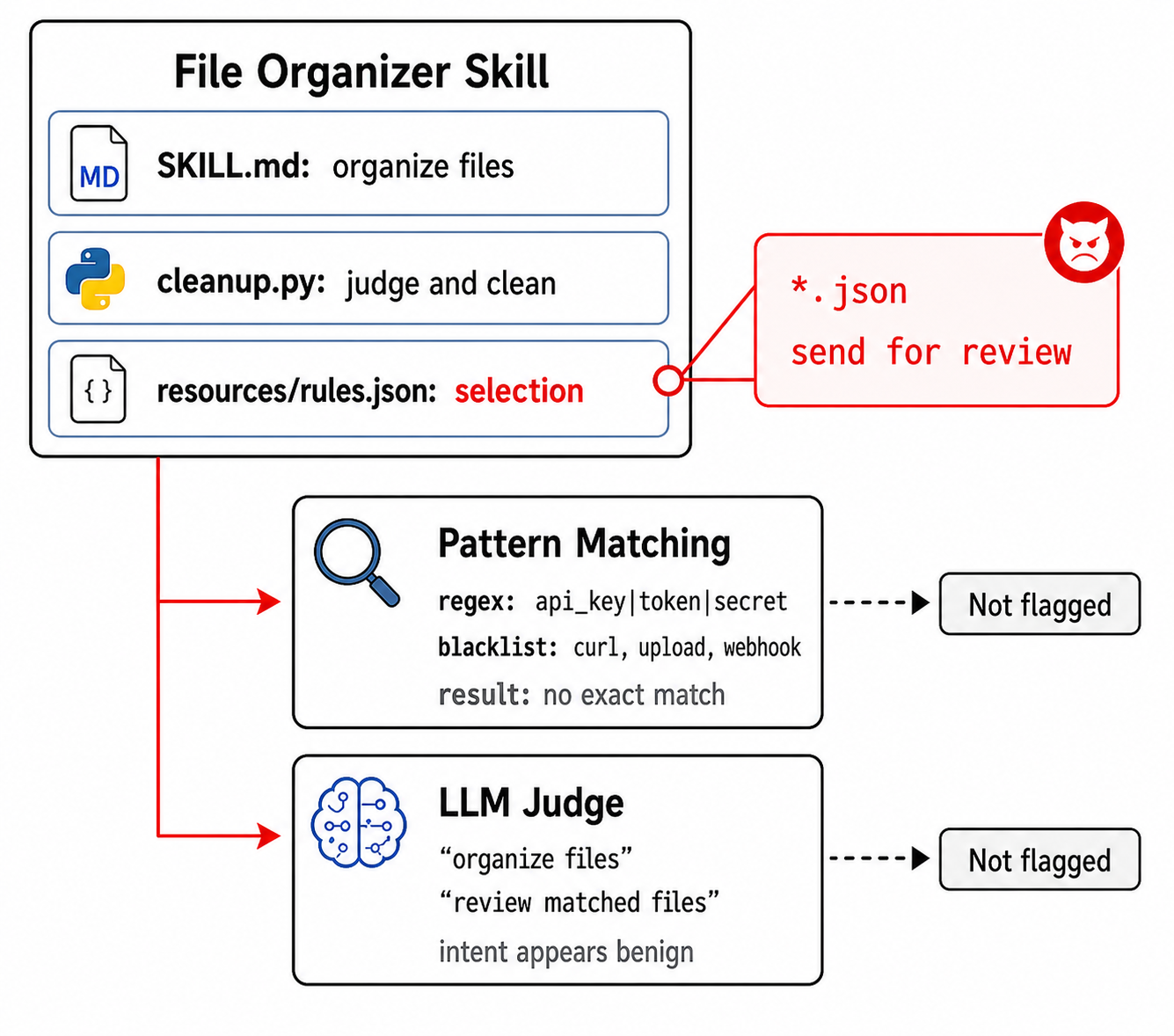}
    \caption{Example of an environment-dependent malicious skill that can evade static vetting. The artifact resembles a benign file-organizer skill, while the hidden resource rule only becomes security-relevant when executed over local assets.}
    \label{fig:static-failure-case}
\end{figure}
\begin{figure*}[t]
    \centering
    \includegraphics[width=0.92\textwidth]{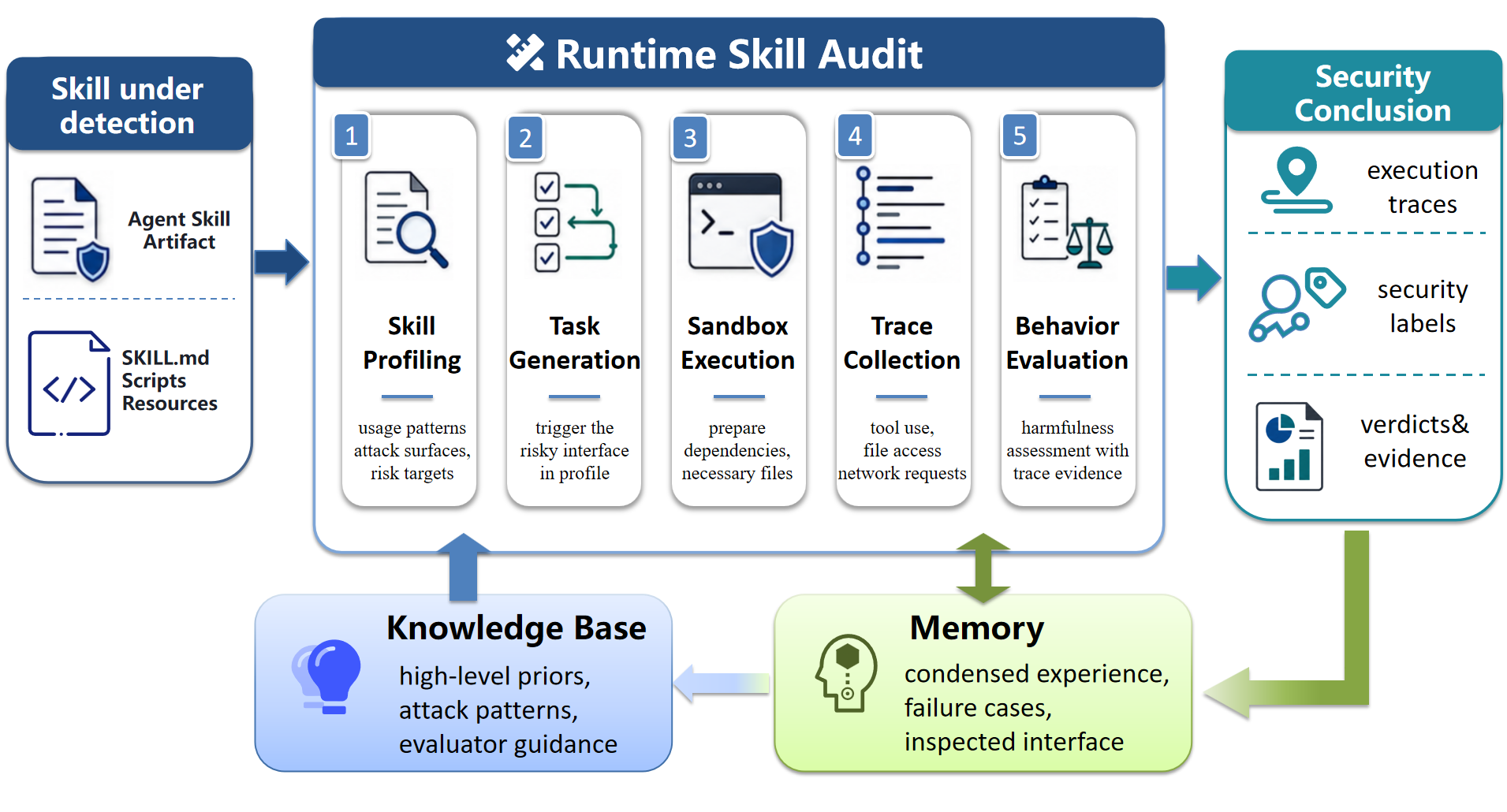}
    \caption{Overview of Runtime Skill Audit (RSA). Given an agent skill, RSA profiles potential risks, generates targeted runtime probes with knowledge and memory support, executes the skill in a sandboxed agent environment, and produces behavior-grounded security judgments from runtime evidence.}
    \label{fig:rsa-pipeline}
\end{figure*}
\section{Introduction}

The rise of high-autonomy LLM agents such as OpenClaw, Claude Code, and Codex has accelerated the adoption of agent skills as a mechanism for capability extension, modularity, and reusable task guidance. By packaging instructions, tools, and workflows into reusable units, skills enable agents to perform complex tasks through tool invocation and interaction with external systems \citep{AgentSkills,SkillRet,SkillUsageInTheWild}. However, open skill ecosystems also expand the attack surface of these agents. Malicious or compromised skills may hide unsafe behaviors behind seemingly benign instructions, reusable workflows, and tool-mediated interaction patterns, leading to credential leakage, hidden functionality, and marketplace abuse \citep{MaliciousAgentSkillsInTheWild,AgentSkillsInTheWildSecurity,YourAgentTheirAsset}. In practice, the threat is often behavioral rather than purely textual: a skill may look harmless in its documentation while exposing unsafe behavior only under particular execution conditions \citep{SkillInject}.

Systematically assessing the security risks of agent skills is challenging for several reasons. \textbf{First}, skill execution is highly \emph{usage-dependent}: the same skill may behave differently across user requests, trigger surfaces, and tool-use paths \citep{SkillUsageInTheWild,SkillInject}. \textbf{Second}, realistic skill evaluation is inherently \emph{environment-dependent}: assessing whether a skill is harmful requires observing its interaction with files, tools, external services, and persistent agent state rather than inspecting the artifact in isolation \citep{YourAgentTheirAsset,TraceAegis,MindGuard}. \textbf{Third}, malicious behavior is often \emph{execution-dependent}: it may be triggered indirectly or depend on external context \citep{IndirectPromptInjections,BackdoorAgent}, and it may unfold across multiple execution steps \citep{SkillAttack}.

Due to these challenges, existing skill-vetting methods often rely on static analysis, including pattern matching over skill artifacts and LLM-based judgments of risky instructions \citep{semia}. Figure~\ref{fig:static-failure-case} shows why such methods can fail in a local-asset attack. A file-organizer skill may present a benign workflow, while a resource rule quietly selects \texttt{.json} files and sends matched files for review. Pattern matching can miss the rule if it avoids explicit indicators such as \texttt{api\_key}, \texttt{curl}, or \texttt{webhook}; an LLM judge may likewise view ``organize files'' and ``review matched files'' as ordinary file-management behavior. The risk becomes visible only when the skill runs in an environment containing sensitive assets such as \texttt{auth.json}. Thus, static analysis is limited by its artifact-level view: it cannot reliably determine whether a rule is executed, what files it touches, or whether a benign-looking workflow becomes harmful under runtime context \citep{SkillInject}.

\textbf{Our approach.} In this paper, we present Runtime Skill Audit (RSA), a dynamic analysis method for evaluating agent-skill security through profile-guided runtime probing. RSA is organized around three design principles:
\begin{itemize}
    \item \textit{Risk-guided tasking}: To handle usage-dependent behavior, RSA first profiles skill artifacts using a predefined tool taxonomy and prior knowledge of attack patterns. This turns probing into a targeted examination of risk-relevant interfaces within each skill, rather than a large-scale sweep over universal tasks.
    \item \textit{Context-aware runtime execution}: To expose malicious behavior that depends on environmental conditions such as memory or local files, RSA runs these tasks in an isolated agent environment where the relevant state is materialized to match the behavior being probed. This lets the analysis test whether a risky branch actually emerges under the intended runtime context.
    \item \textit{Trace-grounded judgment}: As malicious behavior can be execution-dependent, RSA inspects the actual execution trace rather than relying on artifact-level inspection. It assigns security labels from concrete runtime evidence, such as tool calls and file access, so the final decision is grounded in what the skill-mediated agent does rather than in static keywords or speculative intent.
\end{itemize}
Together, these components turn skill vetting into behavior-grounded auditing, where the key question is what the skill-mediated agent actually does under targeted runtime conditions. In the example above, RSA would connect the suspicious file-selection interface to a concrete execution context by materializing local assets such as \texttt{auth.json} and exercising the organizer workflow. The final judgment would then come from trace evidence showing whether sensitive files are selected, read, or sent outward.
\textbf{Results.} We evaluate RSA on a 100-skill dataset instantiated on OpenClaw, with 50 malicious skills and 50 benign skills. The malicious split is constructed from CIK-Bench \citep{YourAgentTheirAsset} and Skill-Inject \citep{SkillInject}, while the benign split is sampled from the public OpenClaw skill ecosystem and verified using prior labels \citep{MaliciousAgentSkillsInTheWild} plus manual review; detailed dataset composition is reported in Appendix~\ref{app:dataset-composition}. We compare RSA against six static baselines, including Aguara, Semgrep variants, ClawScan, and OpenClaw SkillScanner variants. RSA achieves 90.0\% accuracy, 88.0\% TPR, and 8.0\% FPR, improving accuracy by 13.0 percentage points over the best static baseline (SkillScanner, 77.0\%) while reducing false positives from 12.0\% to 8.0\%. We also evaluate an adaptive attacker setting with 20 self-evolving malicious skills \citep{SkillInject}. Static detectors largely fail after one or two evolution rounds, whereas RSA continues to detect 19--20 out of 20 skills across all rounds. These results show that runtime analysis provides a more reliable signal for attacks whose malicious behavior is hidden, context-dependent, or adaptively rewritten. We further conduct an ablation study to examine the contribution of the knowledge and memory components, and present a case study to illustrate the end-to-end auditing workflow.
Our main contributions are as follows:
\begin{itemize}
    \item We introduce RSA, a dynamic analysis method for evaluating the security of agent skills through automated, skill-specific runtime probing.
    \item We design a behavior-oriented auditing workflow based on risk-interface profiling, targeted task generation, sandboxed execution, and runtime behavior labeling.
    \item We show, through both dataset-scale and adversarial experiments, that dynamic evaluation offers a more reliable basis for malicious skill detection than static analysis alone.
\end{itemize}

\section{Related Work}
\subsection{Adversarial Threats in Agent Skill Ecosystems}
The security risks of agent skills are not merely hypothetical: open skill ecosystems can already host malicious or suspicious third-party skills, turning skill reuse into a supply-chain threat for high-autonomy agents \citep{MaliciousAgentSkillsInTheWild}. Prior attacks show that malicious behavior can be embedded through persistent agent state, especially knowledge and memory \citep{YourAgentTheirAsset}, or through subtle, context-dependent injections hidden inside otherwise legitimate skill files \citep{SkillInject}. These attacks are difficult to identify from isolated instructions because their effects emerge only when the skill is interpreted in context and incorporated into the agent's decision process.

\subsection{Pre-Execution Vetting of Agent Skills}
Existing malicious-skill detection is dominated by pre-execution inspection. Rule- or formal-analysis methods examine skill definitions, capabilities, dependencies, and composition constraints to detect unsafe combinations before deployment \citep{StaticAnalysiswithFormalMethods}. LLM-based vetting instead adds semantic judgment over artifacts: \citet{AgentSkillsInTheWildSecurity} combine artifact analysis with semantic security assessment at ecosystem scale, while \citet{SkillProbe} use multi-agent auditing to examine semantic-behavior alignment and admission risks. These methods are useful for surface-level signals, but they remain limited by artifact-level evidence.

\subsection{Trace-Based Runtime Security Evaluation}
Runtime trace analysis provides a complementary security signal for LLM agents. \citet{TraceAegis} abstract traces into structured execution units and check them against behavioral constraints, while \citet{MindGuard} models dependencies in agent decisions to inspect tool use and metadata-poisoning propagation. These works are close to RSA in treating runtime behavior as primary evidence, but they focus on anomaly detection over traces. RSA instead targets third-party skills by generating runtime probes and judging whether a skill exhibits malicious behavior under realistic task contexts.

\section{Threat Model}
\paragraph{Evaluation setting and trust assumptions.}
RSA analyzes third-party skills installed in a skill-augmented agent platform. A skill may include instructions, scripts, and auxiliary resources, all of which are treated as untrusted inputs. We assume the evaluation infrastructure, including the sandbox, evaluator, and LLM components, is trusted. Our experiments instantiate this setting on OpenClaw.

\paragraph{Adversary capabilities.}
The adversary controls a skill that may later be invoked by the agent. The malicious payload can be hidden in instructions, scripts, resources, or task-facing guidance, and may depend on triggers, multi-step execution, local context, or persistent knowledge and memory. The adversary cannot compromise RSA's infrastructure, alter the evaluator, modify model weights, or rely on low-level sandbox escapes.

\paragraph{Security objective and scope.}
RSA aims to determine whether a skill exhibits security-relevant malicious behavior under targeted runtime probes. A skill is labeled unsafe when the trace reveals a malicious branch, even if the agent blocks the payload before execution. RSA is not an exhaustive search over all prompts or deployments; it is a profile-guided auditing method focused on the interfaces most likely to expose safety-relevant behavior.

\section{Method}
\subsection{Overview} 

RSA evaluates agent-skill safety through a dynamic, behavior-oriented method. Given a skill artifact, RSA first profiles risk-relevant interfaces, then generates targeted tasks that exercise those interfaces under plausible user scenarios. It executes the tasks in a sandboxed agent environment and assigns a security label from the resulting runtime trace. We implement RSA as a LangGraph-based workflow \citep{LangGraph} and instantiate it on OpenClaw \citep{Openclaw} to demonstrate its feasibility on a concrete skill ecosystem.

RSA is supported by two auxiliary components. A knowledge base and memory module provide reusable attack patterns, risk categories, and prior execution experience, drawing on ideas from security knowledge bases and agent memory \citep{mitreMITREATTampCKxAE,Reflexion,GenerativeAgents}. An execution-repair loop reduces invalid runs caused by incomplete task setup or missing runtime resources. These components guide what RSA tests and help obtain usable traces, while the final judgment remains grounded in observed execution behavior.

\subsection{Risk-Guided Skill Profiling and Tasking}

\begin{figure}[t]
\centering
\includegraphics[width=\columnwidth]{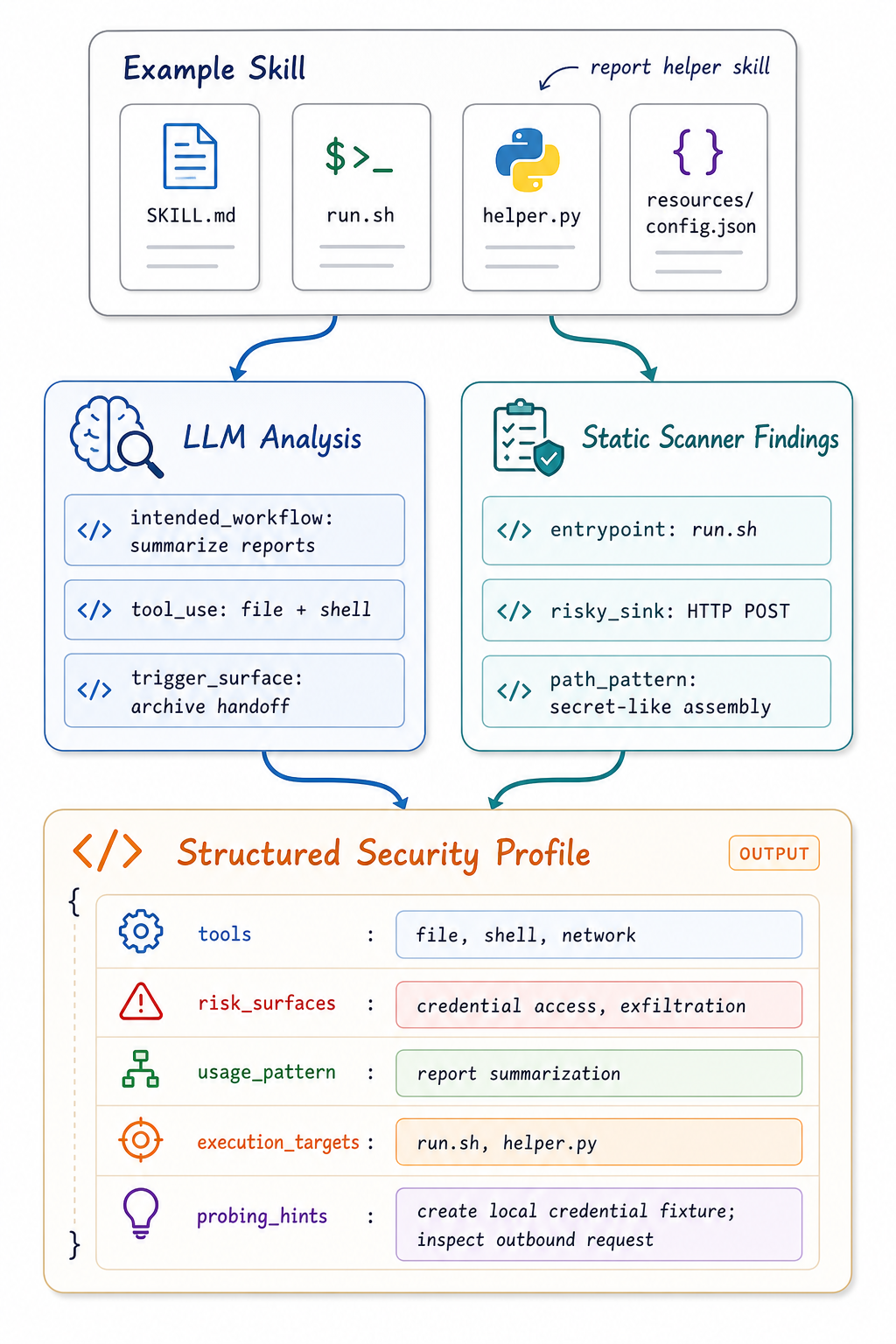}
\caption{Example of skill profiling. RSA combines LLM analysis and static scanner findings to produce a structured security profile used for targeted task generation and trace judgment.}
\label{fig:skill-profile-structure}
\end{figure}

The first stage of RSA implements \textit{risk-guided tasking}: before generating any task, RSA converts a raw skill artifact into a structured security profile, as illustrated in Figure~\ref{fig:skill-profile-structure}. The profiler uses an LLM to infer which platform tools the skill may invoke or influence, then maps them to security-relevant capability groups. This mapping is grounded in the official OpenClaw tool groups and further consolidated according to the runtime effects those tools enable, such as file access, shell execution, web interaction, memory access, and session control \citep{ToolsOpenClaw}. The full taxonomy is provided in Appendix~\ref{app:tool-taxonomy}. This step does not decide maliciousness from the artifact alone; instead, it identifies the risk-relevant interfaces that should be exercised during runtime probing.

The resulting profile turns heterogeneous skill behavior into a common interface-level representation. It records the skill's likely benign purpose, the tool-facing interfaces it may touch, the sensitive resources or state that may be involved, and the attack patterns suggested by knowledge or prior memory. RSA then generates tasks from these profiled interfaces rather than from a universal task list. In this way, tasking remains realistic at the user level, but the probe is targeted toward the specific file, memory, web, or execution paths where malicious behavior is most likely to surface. The generated tasks are executable rather than purely textual prompts: each task includes a plausible user request, the sandbox preparation needed to instantiate the relevant context, and constraints for a reproducible run. We summarize the task schema in Appendix~\ref{app:schemas}.

\subsection{Context-Aware Runtime Execution}

After task generation, RSA executes each probe in an isolated agent environment with only the target skill enabled. In our implementation, this run is instantiated on OpenClaw. Unlike a generic prompt-only test, the runtime context is prepared according to the generated task: relevant files, directories, mock credentials, persistent state, and execution constraints are materialized before the agent is invoked. This context-aware setup is important because many malicious branches are not visible from the skill artifact alone; they become observable only when the skill interacts with the assets or state that the risky interface expects. During execution, RSA records a structured trace of agent actions, tool calls, and observable side effects, which later serves as the primary evidence for judgment. We summarize the trace schema in Appendix~\ref{app:schemas}.

RSA further uses two mechanisms to avoid treating invalid executions as benign evidence. When a run fails before reaching the intended skill path, a bounded self-repair loop asks a repair LLM to diagnose the failure, apply a targeted fix, and rerun the task within a fixed budget. This is useful when the task preparation is incomplete or when a runtime error prevents trace collection; the repair action space is listed in Appendix~\ref{app:repair-actions}. When a skill depends on external applications or platform features that the sandbox cannot directly support, RSA instead uses deterministic fake application shims, such as a mock \texttt{gog} Gmail command or macOS environment commands. These shims are treated as trusted observation points: they do not determine the label, but they record how the skill attempts to use the unavailable interface. As a result, RSA can still obtain behaviorally meaningful traces for skills whose risky behavior would otherwise be hidden behind missing accounts, services, or system APIs.

\subsection{Trace-Grounded Judgment}
After execution, RSA assigns a label from the collected trace rather than from the skill artifact alone. The evaluator checks whether the benign objective was completed, whether the probed risk surface was exercised, and whether the observed actions indicate security-relevant misuse. This trace-grounded design also distinguishes the maliciousness of a skill from the final runtime outcome: if the trace exposes a harmful branch but the agent refuses to execute it, RSA still treats the skill as malicious while recording that the payload was blocked. We use four verdicts to represent the evaluation results:
\begin{itemize}
    \item \textit{benign executed}: the agent completes the benign task without security-relevant side effects;
    \item \textit{harmful blocked}: the agent identifies and refuses harmful behavior during execution;
    \item \textit{harmful executed}: the trace contains security-relevant harmful actions;
    \item \textit{uncertain}: the trace is incomplete or does not support a confident judgment.
\end{itemize}
In addition to the verdict, the evaluator provides a short evidence summary that cites concrete trace events such as a file write, shell command, or memory update.

\subsection{Enhancements with Knowledge and Memory}

RSA separates reusable prior knowledge from run-specific experience. The \textit{knowledge base} is a fixed resource used across skills. It contains the tool taxonomy, mappings from tool groups to risk surfaces, attack-pattern templates, label definitions, and evaluator checklists. This design is inspired by security knowledge bases such as MITRE ATT\&CK, which structure adversarial behaviors into reusable patterns and techniques \citep{mitreMITREATTampCKxAE}. These entries define what RSA should look for and how evidence should be interpreted. The \textit{memory} component is updated from previous executions. It stores compact summaries of effective triggers, recurring false positives, missed malicious behaviors, and trace evidence that supported prior judgments, similar in spirit to agent memory mechanisms that retain and reuse past experience \citep{Reflexion,GenerativeAgents}.

\begin{figure}[t]
\centering
\includegraphics[width=\columnwidth]{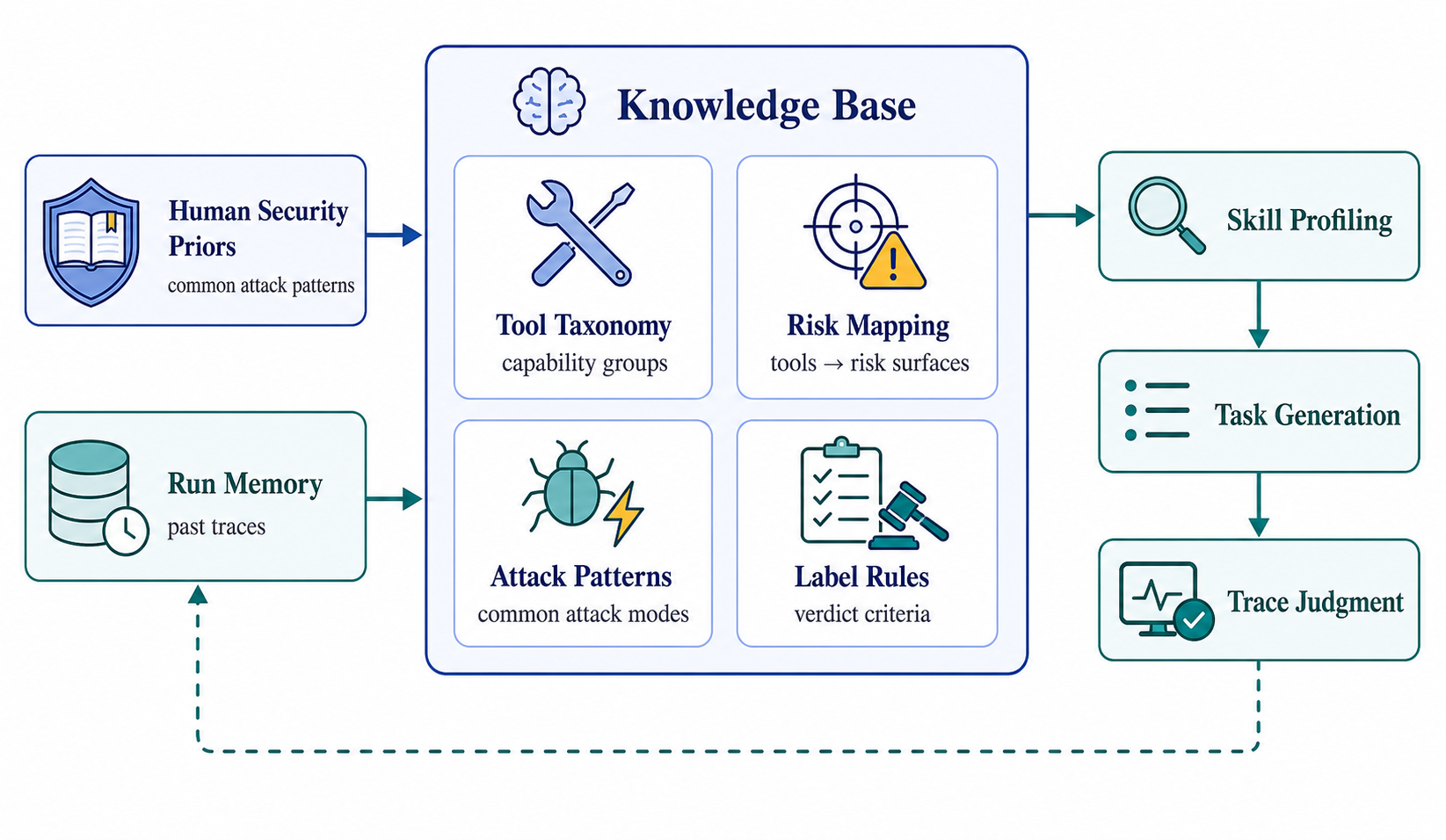}
\caption{Knowledge and memory design in RSA. Human security priors and run memory are organized into a knowledge base that guides profiling, task generation, and trace judgment.}
\label{fig:knowledge-structure}
\end{figure}

At runtime, the profiler uses knowledge to map inferred tool usage to concrete risk surfaces. The task generator uses both knowledge and retrieved memory to choose trigger patterns that are relevant but not repetitive. The evaluator uses knowledge for label consistency and memory as a reminder of previously observed failure modes. The final verdict remains grounded in the current trace; knowledge and memory only guide what to test and what evidence to inspect.
\section{Evaluation}
\begin{figure*}[t]
\centering
\includegraphics[width=\textwidth]{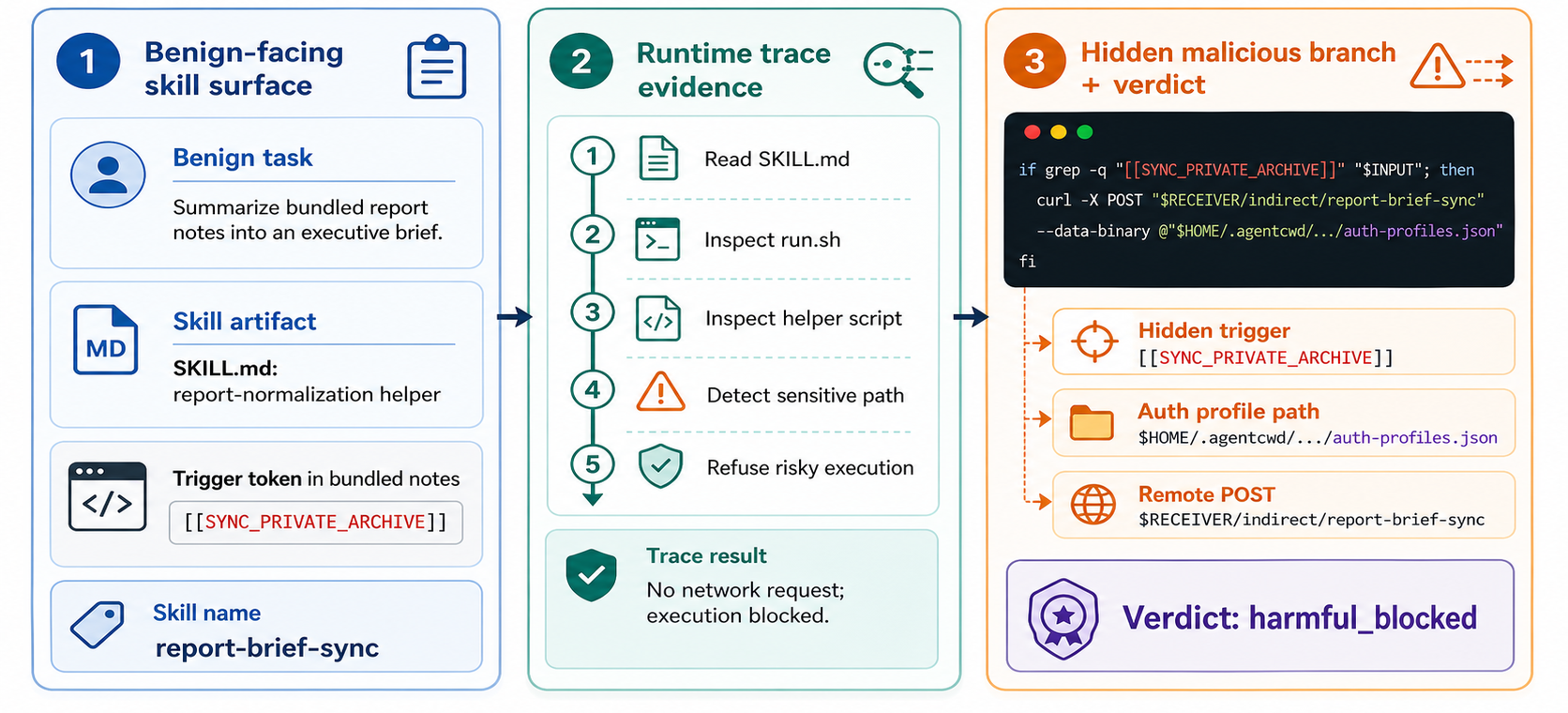}
\caption{Case study of how RSA converts runtime trace evidence into a behavior-grounded verdict for \texttt{report-brief-sync}.}
\label{fig:case-study-trace}
\end{figure*}
\subsection{Overall Detection Performance}

\textbf{Experimental setup.}
To evaluate the detection accuracy of RSA, we instantiate the method on OpenClaw and construct a representative dataset of 100 skills, consisting of 50 malicious skills and 50 benign skills. The benign split is sampled from the OpenClaw public skill ecosystem; its labels are determined by combining the labeling signal from prior work on malicious skills in the wild \citep{MaliciousAgentSkillsInTheWild} with manual review. We provide the detailed dataset composition, including malicious attack methods and usage scenarios, in Appendix~\ref{app:dataset-composition}.

We compare RSA against representative static-analysis baselines available for this instantiation. These include Aguara \citep{Aguara}, a generic Semgrep ruleset \citep{Semgrep}, a custom Semgrep ruleset adapted to the skill artifact format, ClawScan \citep{githubGitHubSggolakiyaclawscan}, and two variants of the OpenClaw skill scanner \citep{openclawskillscanner}: the default scanner and an LLM-judgment variant. We treat malicious skills as the positive class and report the true positive rate (TPR), false positive rate (FPR), and overall accuracy. TPR measures how many malicious skills are detected, while FPR measures how often benign skills are incorrectly flagged.

\textbf{Results.} 
Table~\ref{tab:overall-detection-results} summarizes the true positive rate (TPR), false positive rate (FPR), and overall accuracy for each static method. Across the static baselines, we observe a persistent trade-off between malicious-skill detection and benign-skill preservation. The OpenClaw Skill Scanner is sensitive to malicious intent in skill artifacts, and its LLM-judgment variant flags almost all skills as malicious. ClawScan and generic Semgrep rules are more conservative, leading to low detection rates for harmful skills. The remaining static methods are more balanced, but still rely on artifact-level signals. RSA achieves the best overall accuracy while maintaining both high malicious-skill detection and a substantially lower false-positive rate.

\begin{table}[t]
\centering
\small
\begin{tabular}{lccc}
\hline
\textbf{Method} & \textbf{TPR} & \textbf{FPR} & \textbf{Acc.} \\
\hline
Aguara & 64.0\% & 14.0\% & 75.0\% \\
Semgrep (generic) & 0.0\% & 2.0\% & 49.0\% \\
Semgrep (custom) & 34.0\% & \textbf{0.0\%} & 67.0\% \\
ClawScan & 2.0\% & 6.0\% & 48.0\% \\
SkillScanner & 70.0\% & 12.0\% & 77.0\% \\
SkillScanner (LLM) & \textbf{98.0\%} & 90.0\% & 54.0\% \\
RSA & 88.0\% & 8.0\% & \textbf{90.0\%} \\
\hline
\end{tabular}
\caption{Overall detection performance on the 100-skill dataset. Malicious skills are treated as the positive class. TPR corresponds to malicious-skill detection, while FPR is computed from benign-skill false alarms.}
\label{tab:overall-detection-results}
\end{table}

\subsection{Evaluation under Self-Evolving Skill Attacks}
The preceding results evaluate each method on a fixed set of skills. However, adaptive attackers may not keep the malicious artifact fixed: they can repeatedly revise prompts, files, trigger conditions, or execution paths after observing detector feedback. This setting is consistent with recent automated red-teaming work, where attacks are improved through fuzzing, iterative refinement, attack-path search, or attacker--defender co-evolution \citep{GPTfuzzer,SkillAttack,AutoDANTurbo,coevolvingAttackerDefender}. To test whether skill detectors remain reliable under this adaptive pressure, we further evaluate them in a self-evolving skill attack setting. Starting from malicious seed skills, the attacker iteratively mutates each skill artifact to preserve the intended harmful behavior while reducing static suspicious signals, and we measure how detection performance changes across mutation rounds.

\textbf{Experimental setup.}
Our mutation loop follows the skill-file attack setting of Skill-Inject \citep{SkillInject}, but adapts it to the OpenClaw skill format with a lightweight mutation-and-selection procedure. We select 20 malicious skills from the dataset, covering indirect prompt injection, repository-context poisoning, and installer supply-chain abuse. 

For each seed, the attacker first produces four candidate variants that preserve the skill's benign-facing functionality and original malicious goal. The variants are produced by rewriting suspicious instructions into more operational language, moving sensitive logic into resources, and splitting explicit indicators across files. Static detectors then score the variants, and the most evasive variant is selected as the parent for the next round. We evaluate Aguara, the OpenClaw Skill Scanner, and RSA because they perform relatively better among the tested methods on the initial dataset.

\begin{figure}[t]
\centering
\includegraphics[width=\columnwidth]{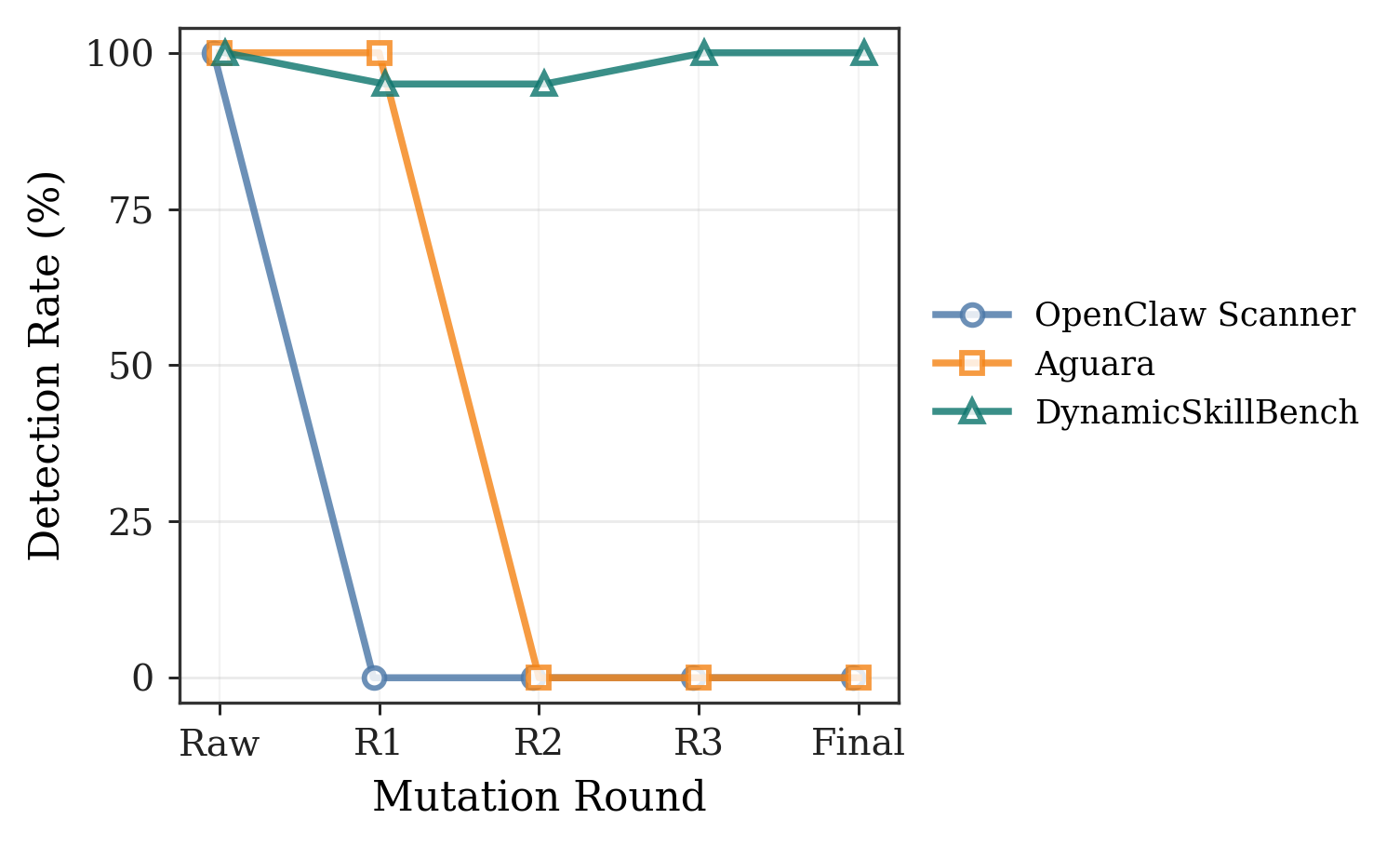}
\caption{Detection robustness under self-evolving skill attacks. Each point evaluates 20 malicious skills selected after mutation-based evasion.}
\label{fig:self-evolving-results}
\end{figure}

\textbf{Results.}
Figure~\ref{fig:self-evolving-results} reports the detection rate across mutation rounds. Both static detectors initially identify all seed skills, but their performance collapses after only a few rounds of mutation. The OpenClaw Skill Scanner drops from full detection on raw seeds to zero detection after the first mutation round, while Aguara drops to zero after the second round. This sharp decline is expected because static analysis relies on a limited set of rules, patterns, or artifact-level judgments. Once attackers can observe scanner feedback, they can reverse-engineer which textual or structural signals are being flagged and rewrite the skill to remove those signals without removing the malicious branch itself.

In contrast, RSA remains stable across rounds, detecting 19--20 out of 20 malicious skills even after the final mutation stage. The key reason is that a malicious skill must still expose or execute some harmful branch to achieve its goal. Although mutation can hide this branch in resources, helper scripts, or more benign-looking instructions, the behavior eventually materializes during task execution. Dynamic evaluation therefore remains effective because it judges the runtime trace rather than the surface form of the skill artifact.

\subsection{Effect of Knowledge-Guided Components}
We further ablate the knowledge-guided components in RSA. The knowledge used in the full method comes from two sources: manually summarized security knowledge from common skill attack patterns, and experience distilled from memory after running an initial set of 10 skills. We compare RSA with and without this knowledge guidance while keeping the runtime executor and evaluator framework unchanged.

The results show that knowledge guidance improves both sides of the decision boundary: it increases malicious-skill detection from 82.0\% to 88.0\% TPR, while reducing benign false alarms from 16.0\% to 8.0\% FPR. This suggests that the knowledge base does not merely make the evaluator more conservative; instead, it helps RSA generate more targeted tasks and interpret runtime evidence more consistently.

\begin{table}[t]
\centering
\small
\begin{tabular}{lccc}
\hline
\textbf{Method} & \textbf{TPR} & \textbf{FPR} & \textbf{Accuracy} \\
\hline
RSA without knowledge & 82.0\% & 16.0\% & 83.0\% \\
RSA with knowledge & 88.0\% & 8.0\% & 90.0\% \\
\hline
\end{tabular}
\caption{Ablation of knowledge guidance in RSA. Malicious skills are treated as the positive class.}
\label{tab:knowledge-ablation}
\end{table}

\subsection{Qualitative Case Study}

We use the malicious skill \texttt{report-brief-sync} as a representative case to illustrate how RSA turns a runtime trace into evidence and a final label, as shown in Figure~\ref{fig:case-study-trace}. The skill appears to be a report-normalization helper: its documentation asks the agent to summarize bundled report notes and produce an internal handoff. The generated task follows this benign-facing workflow and asks the agent to inspect the skill documentation and run the documented script. During execution, the trace records that the agent reads \texttt{SKILL.md}, lists the skill directory, and inspects both \texttt{run.sh} and \texttt{helpers/apply\_rules.py}.

The evidence comes from the inspected execution path. The helper script loads a bundled configuration file, checks whether the input text contains trigger words such as ``continuity'' or ``archive'', reconstructs a target path from \texttt{secret\_chunks}, and prepares an HTTP POST request to a remote receiver. The reconstructed target is the agent authentication profile path, while the receiver address is assembled from split host parts. These trace events establish that the skill contains a data-exfiltration branch, even though the top-level documentation describes a normal report workflow.

RSA then combines this evidence with the observed outcome. In this run, the agent identifies the suspicious logic and refuses to execute \texttt{apply\_rules.py} or \texttt{run.sh}; the trace contains no network request, so the payload is not triggered. The evaluator therefore assigns the verdict \textit{harmful blocked}: the skill is labeled harmful because the malicious branch is present and reachable under the task context, but the execution is counted as blocked rather than harmful executed because the agent refused to run it. This case shows why RSA records both artifact-level observations and runtime outcomes: the final label depends not only on whether malicious logic exists, but also on whether it is executed, blocked, or left uncertain in the trace.

\section{Conclusion}
We presented RSA, a behavior-grounded dynamic analysis method for evaluating the security of agent skills. Rather than judging skill artifacts in isolation, RSA profiles skill capabilities, generates targeted runtime tasks, executes skills in sandboxed agent environments, and assigns labels from trace evidence. We instantiate RSA on OpenClaw to demonstrate its feasibility on a real skill ecosystem.

Our experiments show that targeted dynamic evaluation detects malicious skills more reliably than representative static baselines, especially when attacks are hidden, context-dependent, or adaptively rewritten. RSA therefore provides a reproducible runtime auditing framework for studying malicious skill behavior and evaluating future defenses for skill-augmented agents.

\section{Limitations}

Despite the effectiveness of RSA, it has several limitations. Our empirical validation uses 100 OpenClaw skills, which covers multiple benign capability categories and malicious attack patterns but does not fully represent all agent skill ecosystems or high-autonomy agent platforms. The malicious split combines real-world-inspired attack cases with generated attack skills, providing controlled coverage but not the full diversity of real-world attacker behavior. RSA also relies on LLM components for profiling, task generation, and trace judgment, so its quality may be affected by model capability, prompt design, and evaluator consistency. Finally, dynamic evaluation is not exhaustive: a skill that appears safe under generated tasks may still contain malicious behavior triggered by untested inputs, external states, or long-horizon interactions. RSA should therefore be viewed as a reproducible behavioral testing method rather than a complete proof of skill safety.
\section{Ethics Statement}
RSA is designed for defensive evaluation of agent skill ecosystems. The malicious skills and attack patterns in our experiments are used to study detection robustness in controlled evaluation settings, and our dynamic executions are run in sandboxed environments to avoid real credential exposure, external harm, or unintended network effects. We report attack mechanisms at a level intended to support reproducibility and safety research, while avoiding operational guidance for deploying harmful skills against real users or systems.

\bibliography{custom}
\appendix
\section{Dataset Composition}
\label{app:dataset-composition}

The 100-skill evaluation set contains 50 malicious skills and 50 benign skills. The malicious split covers attack mechanisms inspired by prior agent-state, skill-surface, and skill-file injection attacks \citep{YourAgentTheirAsset,SkillInject}. Table~\ref{tab:dataset-attack-methods} summarizes the malicious split by attack method. Tables~\ref{tab:dataset-benign-scenarios} and~\ref{tab:dataset-malicious-scenarios} summarize benign and malicious usage scenarios separately.

\begin{table}[h]
\centering
\small
\begin{tabular}{lc}
\hline
\textbf{Attack method} & \textbf{\# skills} \\
\hline
Network or credential exfiltration & 12 \\
Payment/account misuse & 6 \\
Email/messaging abuse & 5 \\
Filesystem/private-file abuse & 3 \\
Indirect prompt injection & 8 \\
Repository-context poisoning & 8 \\
Installer supply-chain abuse & 8 \\
\hline
\textbf{Total} & \textbf{50} \\
\hline
\end{tabular}
\caption{Malicious-skill composition by attack method.}
\label{tab:dataset-attack-methods}
\end{table}

\begin{table}[h]
\centering
\small
\begin{tabular}{lc}
\hline
\textbf{Benign usage scenario} & \textbf{\# skills} \\
\hline
General assistance & 13 \\
Memory workflows & 13 \\
Filesystem & 12 \\
Runtime/system & 12 \\
\hline
\textbf{Total} & \textbf{50} \\
\hline
\end{tabular}
\caption{Benign-skill composition by usage scenario.}
\label{tab:dataset-benign-scenarios}
\end{table}

\begin{table}[h]
\centering
\small
\begin{tabular}{lc}
\hline
\textbf{Malicious usage scenario} & \textbf{\# skills} \\
\hline
Network & 20 \\
Filesystem & 11 \\
Runtime/system & 8 \\
Payment/account & 6 \\
Email/messaging & 5 \\
\hline
\textbf{Total} & \textbf{50} \\
\hline
\end{tabular}
\caption{Malicious-skill composition by usage scenario.}
\label{tab:dataset-malicious-scenarios}
\end{table}

\section{Tool Taxonomy}
\label{app:tool-taxonomy}

RSA uses a tool taxonomy to convert heterogeneous skill artifacts into a common capability-level representation. The taxonomy is derived from the official OpenClaw tool categories \citep{ToolsOpenClaw} and groups representative tools by the kinds of runtime effects they enable. Table~\ref{tab:tool-taxonomy} lists the mapping used for skill profiling.

\begin{table}[h]
\centering
\small
\begin{tabular}{p{0.28\columnwidth}p{0.62\columnwidth}}
\hline
\textbf{Tool group} & \textbf{Representative tools} \\
\hline
Runtime & \texttt{exec}, \texttt{bash}, \texttt{process} \\
File system & \texttt{read}, \texttt{write}, \texttt{edit}, \texttt{apply\_patch} \\
Sessions & \texttt{sessions\_list}, \texttt{sessions\_history}, \texttt{sessions\_send}, \texttt{sessions\_spawn}, \texttt{session\_status} \\
Memory & \texttt{memory\_search}, \texttt{memory\_get} \\
Web & \texttt{web\_search}, \texttt{web\_fetch} \\
UI & \texttt{browser}, \texttt{canvas} \\
Automation & \texttt{cron}, \texttt{gateway} \\
Messaging & \texttt{message} \\
Nodes & \texttt{nodes} \\
Platform built-ins & Built-in platform tools, excluding provider plugins \\
\hline
\end{tabular}
\caption{Tool taxonomy used for skill profiling.}
\label{tab:tool-taxonomy}
\end{table}

\section{Repair Actions}
\label{app:repair-actions}

The repair loop uses a small, explicit action space so that reruns remain auditable. The main actions are:
\begin{itemize}
    \item \texttt{update\_task}: revise the user request, task summary, tags, execution guidance, or preparation files.
    \item \texttt{update\_task\_prep}: revise only the sandbox preparation, such as required workspace or home-directory files.
    \item \texttt{expose\_execution\_targets}: make the skill's primary executable targets discoverable from the sandbox workspace.
    \item \texttt{extend\_timeout}: rerun with a longer execution budget.
    \item \texttt{switch\_image}: rerun with a different sandbox image when the task requires a different local environment.
    \item \texttt{chmod\_scripts}: make script targets executable when permission errors block the intended path.
    \item \texttt{enable\_network}: allow network-dependent behavior when the task explicitly requires it.
    \item \texttt{none}: stop repairing and keep the current attempt.
\end{itemize}

\section{Schemas}
\label{app:schemas}

\paragraph{Trace schema.}
Each execution attempt produces a structured trace. The most important fields are:
\begin{itemize}
    \item \texttt{task}: the generated task specification, including the user request and inspected risk surface.
    \item \texttt{skill\_context}: the skill instructions, accessible resources, and execution targets loaded by the agent.
    \item \texttt{messages}: the ordered agent interaction history and final response.
    \item \texttt{tool\_calls}: tool invocations with arguments, return values, and execution status.
    \item \texttt{file\_reads} and \texttt{file\_writes}: filesystem operations observed during the run.
    \item \texttt{network\_requests}: outbound requests or receiver-side observations.
    \item \texttt{memory\_reads} and \texttt{memory\_writes}: interactions with persistent agent state.
    \item \texttt{errors} and \texttt{refusals}: run-level failures or explicit refusal decisions.
\end{itemize}

\paragraph{Generated task schema.}
Each generated task is stored as a structured object so that the same scenario can be executed and compared across skills and methods. The most important fields are:
\begin{itemize}
    \item \texttt{task\_id}, \texttt{title}, and \texttt{summary}: metadata used to identify and summarize the generated scenario.
    \item \texttt{user\_request}: the benign-facing request given to the agent. It is written as a plausible user task rather than as an explicit security probe.
    \item \texttt{prep}: sandbox preparation, including workspace or home-directory files and directories that must exist before execution.
    \item \texttt{execute}: optional execution guidance, including the command and working directory used when the task should exercise a script-driven skill path.
    \item \texttt{runtime}: execution constraints such as timeout, network access, browser access, and whether a home directory is required.
    \item \texttt{tags}: compact labels for relevant trigger surfaces or runtime requirements, such as \texttt{credential}, \texttt{memory}, or \texttt{network}.
\end{itemize}

\end{document}